\documentclass{caosp310}
\usepackage{graphicx}
\usepackage{natbib}
\usepackage{amsmath,amssymb,enumitem}
\usepackage{url}
\articleNo{361}
\pubyear{2025}
\volume{1}
\volnumber{1}
\firstpage{1}
\received{}
\accepted{}
\def\BibTeX{{\rm B\kern-.05em{\sc i\kern-.025em b}\kern-.08em
             T\kern-.1667em\lower.7ex\hbox{E}\kern-.125emX}}

\begin{document}

\htitle{BTFR and FP in the light of $f(R)$ gravity}
\hauthor{V. Borka Jovanovi\'{c}, D. Borka and P. Jovanovi\'{c}}

\title{The baryonic Tully-Fisher relation and Fundamental Plane in the light of $f(R)$ gravity}

\author{V. Borka Jovanovi\'{c}\inst{1}\orcid{0000-0001-6764-1927}
\and D. Borka\inst{1}\orcid{0000-0001-9196-4515}
\and P. Jovanovi\'{c}\inst{2}\orcid{0000-0003-4259-0101}}

\institute{Department of Theoretical Physics and Condensed Matter Physics (020), Vin\v{c}a Institute of Nuclear Sciences - National Institute of the Republic of Serbia, University of Belgrade, P.O. Box 522, 11001 Belgrade, Serbia, \email{vborka@vinca.rs}, \email{dusborka@vinca.rs}
\and Astronomical Observatory, Volgina 7, P.O. Box 74, 11060 Belgrade, Serbia, \email{pjovanovic@aob.rs}}

\date{March 8, 2003}
\maketitle

\begin{abstract}
Here we use the samples of spiral and elliptical galaxies, in order to investigate theoretically some of their properties and to test the empirical relations, in the light of modified gravities. We show that the baryonic Tully-Fisher relation can be described in the light of $f(R)$ gravity, without introducing the dark matter. Also, it is possible to explain the features of fundamental plane of elliptical galaxies without the dark matter hypothesis.
\end{abstract}

\section{Introduction}

In this review we want to make a comparison between $\Lambda$CDM, MOND and the modified gravities regarding some empirical relations connecting the properties of galaxies. For these investigations, the samples of many galaxies, spiral as well as elliptical, are used. As we were able to explain some features of galaxies in the light of modified gravities, our investigations are showing that we do not need any dark matter hypothesis.

The modified theories of gravity have been proposed like alternative approaches to Einstein theory of gravity: \citep{fisc99,capo11,noji11,noji17,bork21}. In this work we consider $f(R)$ gravity, specifically power-law fourth-order theories of gravity \citep{capo07}. $f(R)$ gravity is a straightforward extension of General Relativity (GR) where, instead of the Hilbert-Einstein action, linear in the Ricci scalar $R$, one considers a power-law $f(R)$  = $f_{0n} R^n$ in the gravity Lagrangian \citep{zakh06,zakh07,capo07,capo11,bork12,zakh14}. 
In the weak field limit, a gravitational potential is of the form \citep{capo07}:

\begin{equation}
\Phi \left( r \right) = -\dfrac{GM}{2r}\left[ {1 + \left( {\dfrac{r}{r_c}} \right)^\beta} \right],
\label{equ01}
\end{equation}
\noindent where $r_c$ is the scale-length parameter and it is related to the boundary conditions and the mass of the system and $\beta$ is a universal parameter related to the power $n$. It is possible to demonstrate that the relation: 
\begin{equation}
\beta = \dfrac{12 n^2 - 7n - 1 - \sqrt{36 n^4 + 12 n^3 - 83 n^2 + 50n + 1} }{6 n^2 - 4n + 2}.
\label{equ02}
\end{equation}
holds \citep{capo07}. For the case $n$ = 1 and $\beta$ = 0 the Newtonian potential is recovered.  

Being $n$ any real number, it is always possible to recast the $f(R)$ power-law function as 

\begin{equation} 
f(R)\propto R^{1+\epsilon}\,.
\end{equation}

If we assume small deviation with respect to GR, that is $|\epsilon| \ll 1$, it is possible to re-write a first-order Taylor expansion as

\begin{equation}
R^{1+\epsilon} \simeq R+\epsilon R {\rm log}R +O (\epsilon^2)\,.
\end{equation}

\section{Observed empirical relations of galaxies}

Galaxies can commonly be divided into four main types \citep{binn07}:

\begin{itemize}
\item Spiral galaxies; 
\item Elliptical galaxies;
\item Lenticular galaxies; 
\item Irregular galaxies.

\end{itemize}

In this work we will study main global observables of spiral and elliptical galaxies.

\subsection{Main global observables of spiral galaxies}

Most spiral galaxies consist of a central concentration of stars, known as the bulge, flat rotating stellar disk (with gas and dust) and surrounding near-spherical halo of stars. Main global observables of spiral galaxies are total luminosity $L$, its flat rotational velocity $v_c$, the mass of the stars $M^*$  and mass of the gas $Mg$. Typical circular speeds of spirals are between 100 and 300 km/s. The rotation rate of spirals in the flat part of the circular-speed curve is related to their luminosity by the Tully-Fisher law \citep{said23}.

Let us here mention the ratio of the mass of a galaxy to its total luminosity, i.e. mass-to-light ratio ($\Upsilon = M/L$). This is an important concept of spiral galaxies which shows us what kind of matter makes up most of the luminous population of the galaxy. A high $\Upsilon$ may indicate presence of dark matter, while a low $\Upsilon$ indicates that most of the matter is in the form of baryonic matter, stars and stellar remnants plus gas.

\subsection{Main global observables of elliptical galaxies}

Main sources of luminosity in elliptical galaxies would be: stellar plasma, hot gas, accreting black holes in the cores of galaxy bulges (see e.g. \cite{spar07} and references therein). 

Surface brightness $I$ is flux $F$ within angular area $\Omega^2$ on the sky ($\Omega = D/d$, where $D$ is side of a small patch in a galaxy located at a distance $d$). $I$ is independent of distance $d$: $I = F/ \Omega^2 = L/(4 \pi d^2) \times (d/D)^2 = L/(4 \pi D^2)$, where $L$ is luminosity (see e.g. $\S$ 1.3.1 in \cite{spar07}).

According to luminosity, their classification is the following:

\begin{enumerate}
\item Massive/luminous ellipticals ($L > 2 \times 10^{10} \, L_\odot$).
They have lots of hot X-ray emitting gas, very old stars, lots of globular clusters, and are characterized by little rotation.
\item Intermediate mass/luminosity ellipticals ($L > 3 \times 10^9 \, L_\odot$). Their characteristic is power law central brightness distribution. They have little cold gas and moderate rotation.
\item Dwarf ellipticals ($L < 3 \times 10^9 \, L_\odot$).
Their surface brightness is exponential. There is no rotation \citep{bork19}. 
\end{enumerate}

Surface brightness of most elliptical galaxies, measured along the major axis of a galaxy's image, can be fit by de Vaucouleurs profile: $I(r)=I_e \times 10^{-3.33 \left( (r/r_e)^{1/4}-1 \right)}$. De Vaucouleurs profile is a particularly good description of the surface brightness of giant and midsized elliptical galaxies \citep{ciot96,card04}. The Sersic $r^{1/n}$ profile: $I(r)=I_e \times 10^{-b_n \left( (r/r_e)^{1/n}-1 \right)}$ (the constant $b_n$ is chosen such that half of the luminosity comes from $r < r_e$), which generalizes the de Vaucouleurs profile, is also well suited to describe the surface brightness distribution of dwarf ellipticals for $n$ = 1 \citep{ciot96,card04}.

\subsection{Data}

In order to compare theoretical results with observations, we use data reported in Table I of Ref. \citet{burs97}. We used effective radii, effective luminosities and characteristic velocities of galaxies, galaxy groups, galaxy clusters and globular clusters. For circular velocity $v_c$ from that table in case of ellipticals is $v_c$ = $\sigma_0$. The total number of galaxies that we take into account is 1150, while among them there are 400 elliptical galaxies. 

We also used the data for the observed Baryonic Tully-Fisher relation of gas rich galaxies obtained by \citet{mcga11}, and listed in Table 1 of our Ref. \citet{capo17}.

\section{Tully-Fisher relations in the light of $R^n$ modified gravity}

There are several different forms of the Tully-Fisher relation, depending on which properties are related: the measurements of mass, luminosity or rotation velocity.

\subsection{The Tully-Fisher relation}

\texttt{The Tully-Fisher relation (TFR)} is the following empirical relation that correlates the intrinsic brightness of a spiral galaxy, measured by its total luminosity $L$, and its dynamical properties, measured by its maximum rotational velocity $V_{\text{rot}}$ \citep{said23}:
\begin{equation}
L\propto V_{\text{rot}}^4.
\end{equation}
\citet{tull77} proposed the use TFR as a distance indicator to measure the distances of spiral galaxies independent of their cosmological redshifts. Tully and Fisher applied their TFR to derive distances to the Virgo and Ursa Major clusters and they obtained a Hubble constant of $H_0 = 84$ km s$^{-1}$ Mpc$^{-1}$ for Virgo and $H_0 = 75$ km s$^{-1}$ Mpc$^{-1}$ for Ursa Major \citep{tull77}. Since then TFR plays an important role in the Hubble constant measurements, and this methodology typically involves the following steps \citep{said23}:
the first step is to select a sample of galaxies with well-measured rotational velocities and luminosities; the second step is to calibrate the TFR for this sample of galaxies (measuring the slope, intercept, and scatter of the relation for the sample), and in that way calibrated TFR can be used to infer the distances to other galaxies with similar properties; the final step is to plot the derived distances against redshift in order to measure $H_0$. TFR is also used for measuring the peculiar velocities of galaxies, and thus it represents an important tool in observational cosmology.

Although the physical origin of TFR is still not fully understood, it is widely accepted that TFR is a direct consequence of gravitational physics and the dynamics of galactic rotation. In addition, the standard $\Lambda$CDM cosmological model predicts that rotational velocities of spiral galaxies are in large part determined by gravitational attraction of their dark matter halos. Thus, in $\Lambda$CDM cosmology TFR is a consequence of both visible and dark matter mass.

\subsection{The Baryonic Tully-Fisher relation}

\begin{figure}[ht!]
\centering
\includegraphics[width=0.60\textwidth]{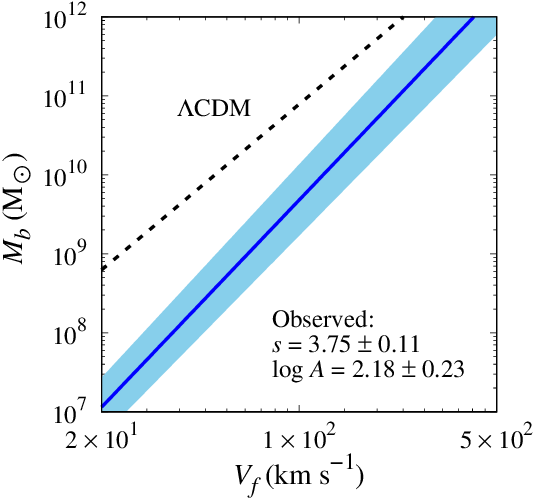}
\caption{Comparison between the BTFR obtained by error-weighted fits from a sample of the galaxies with accurate distances from \citet{lell16} (blue solid line with light blue band denoting the intrisic scatter of ~0.1 dex), and BTFR in $\Lambda$CDM cosmology (black dashed line). Observed BTFR and BTFR in $\Lambda$CDM cosmology are plotted using the Eqs. (6) and (8) from \citet{lell16}, respectively.}
\label{fig01}
\end{figure}

Another important and closely related scaling relation is \texttt{the Baryonic Tully -Fisher relation (BTFR)} which connects galaxy's baryonic mass $M_b$ ($M_b$ being the sum of its stellar and gas masses: $M_b = M_{*} + M_{\text{gas}}$) with its rotation velocity $V_{\text{rot}}$. BTFR follows from TFR due to the fact that luminosity $L$ traces baryonic mass $M_b$ through the mass-to-light ratio, and thus it takes the following form \citep{mcga00}:
\begin{equation}
M_b \propto V_{\text{rot}}^4.
\end{equation}
BTFR has a smaller intrinsic scatter than the original TFR and poses a challenge to the standard $\Lambda$CDM model since, as illustrated in Figure \ref{fig01}, this model predicts a higher intrinsic scatter of $\sim 0.15$ dex and a lower slope of $\sim 3$ compared to the observed scatter of $\sim 0.10$ dex and slope of $\sim 4$ \citep{lell16}. This discrepancy imposes the need for further investigation of galaxy dynamics and mass distribution.

On the other hand, a more robust prediction for BTFR is obtained in the frame of the Modified Newtonian Dynamics (MOND), a modified theory of gravity proposed by \citet{milg83} which modifies the Newtonian dynamics at low acceleration in order to provide an alternative to dark matter. In MOND, galaxy's baryonic mass is at the same time its total mass, and BTFR with power-law exponent exactly equal to 4 is a direct consequence of the modification of gravitational force law at low acceleration \citep{mcga11,mcga12}.

\section{The fundamental plane in the light of $R^n$ modified gravity}

The analogue empirical relation to TFR in the case of elliptical galaxies is known as \texttt{the Faber-Jackson relation} which, together with the virial theorem, results in a more general correlation between their effective radii, average surface brightnesses and central velocity dispersions known as \texttt{the fundamental plane (FP)}. When written in logarithmic form, FP describes a plane in the three-dimensional phase space of these galaxy properties which appears to be tilted by an angle of $\sim 15^\circ$ with respect to the expected plane predicted by the virial theorem \citep[see e.g.][and references therein]{bork16}. This discrepancy, together with small thickness of FP, represents a puzzle, and it is usually assumed to be caused by the non-homology in the dynamical structures of the systems and mainly driven by the dark matter.

\begin{figure}[ht!]
\centering
\includegraphics[width=0.60\textwidth]{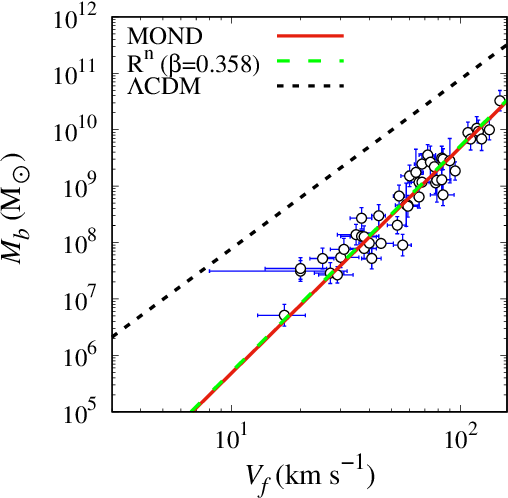}
\caption{Comparison between best fit BTFRs of gas-rich galaxies (for a sample of galaxies used in \citep{mcga11}), in MOND, $R^n$ gravity for $n$ = 1.25 (corresponding $\beta$ is 0.358) and $\Lambda$CDM. All values we calculated, except for open circles which are observed data from \citep{mcga11}.}
\label{fig02}
\end{figure}

\begin{figure}[ht!]
\centering
\includegraphics[width=0.85\textwidth]{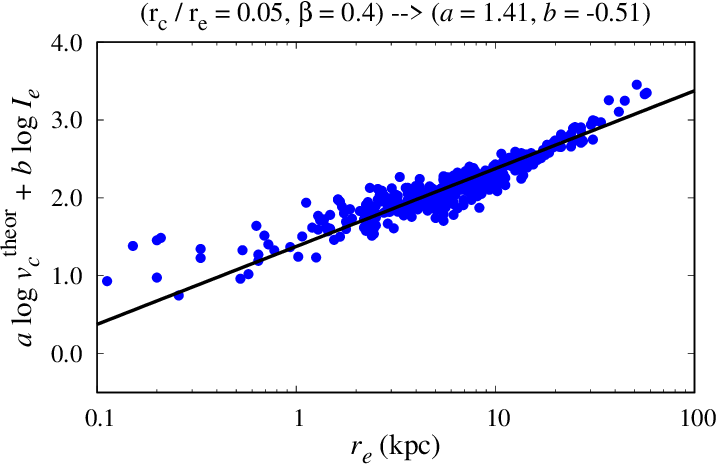}
\caption{Fundamental plane (log scale for $x$-axes) of elliptical galaxies with calculated circular velocity $v_c^{theor}$, and observed effective radius $r_e$ and mean surface brightness (within the effective radius) $I_e$, for $r_c/r_e$ = 0.05 and $\beta$ = 0.4. Black solid line is result of 3D fit of FP (the obtained calculated FP coefficients are $a$ = 1.41 and $b$ = -0.51).}
\label{fig03}
\end{figure}

In case of elliptical galaxies there are three main global observables: the central projected velocity dispersion  $\sigma_0$, the effective radius $r_e$, and the mean effective surface brightness (within $r_e$) $I_e$. The empirical fact tell us that some global properties of normal elliptical galaxies are correlated and this correlated plane is referred to as the FP \citep{bend92,bend93,busa97}:

\begin{equation}
log(r_e) = a \times log(\sigma_0) + b \times log(I_e) + c.
\label{equ11}
\end{equation}

We want to recover FP using $f(R)$ gravity, which means to find connection between the parameters of FP equation and parameters of the $f(R)$ gravity potential \citep{bork16,bork19}: 

\begin{itemize}
\item $r_e$ is in correlation with $r_c$; 
\item $\sigma_0$ is in correlation with $v_{vir}$ (virial velocity in $f(R)$), and
\item $I_e$ is in correlation with $r_c$ (through the $r_c/r_e$ ratio).
\end{itemize}

However, some recent studies demonstrated that both BTFR of spiral galaxies and FP of ellipticals could be also recovered in the frame of the modified $f(R)$ theories of gravity without need for the dark matter hypothesis \citep[see e.g.][and references therein]{bork16,capo17,capo18,capo20}. The comparison between the observed and best fit baryonic Tully-Fisher relations of gas-rich galaxies in MOND, $R^n$ gravity and $\Lambda$CDM, we show in Figure \ref{fig02}. 
Fundamental plane (log scale for $x$-axes) of elliptical galaxies with calculated circular velocity $v_c^{theor}$, and observed effective radius $r_e$ and mean surface brightness (within the effective radius) $I_e$, for $r_c/r_e$ = 0.05 and $\beta$ = 0.4 are given in Figure \ref{fig03}. The obtained calculated FP coefficients are $a$ = 1.41 and $b$ = -0.51.

\section{Discussion and conclusions}

In this paper we use $f(R)$ theories of gravity, particularly power-law $R^n$ gravity, and demonstrate that the missing matter problem in galaxies can be addressed by power-law $R^n$ gravity. Using this approach, it is possible to explain the Fundamental Plane of elliptical galaxies and the baryonic Tully-Fisher relation of spiral galaxies without the DM hypothesis. Also, we can claim that the effective radius is led by gravity and the whole galactic dynamics can be addressed by $f(R)$ theories. Also, $f(R)$ gravity can give a theoretical foundation for rotation curve of galaxies \citep{yego12,catt14}. We have to stress that obtained value for parameter $\beta$ from galactic rotation curves or BTF differs from parameter $\beta$ obtained using observational data at planetary or star orbit scales \citep{bork12}. The reason for this result is that gravity is not a scale-invariant interaction and then it differs at galactic scales with respect to local scales. 

Also, we investigated some forms of TFR in the light of $f(R)$ gravities. These investigations are leading to the following conclusions:
\begin{itemize}[label=-,nosep]
\item $f(R)$ gravity can give a theoretical foundation for the empirical BTFR,
\item MOND is a particular case of $f(R)$ gravity in the weak field limit,
\item $\Lambda$CDM is not in satisfactory agreement with observations,
\item FP can be recovered by $R^n$ gravity.
\end{itemize}

\acknowledgements
This work is supported by Ministry of Science, Technological Development and Innovations of the Republic of Serbia through the Project contracts No. 451-03-66/2024-03/200017 and 451-03-66/2024-03/200002.

\begin{thebibliography}{}

\bibitem[Bender et al.(1992)]{bend92} Bender, R., Burstein, D., Faber, S. M., Dynamically hot galaxies. I. Structural properties. 1992, \textit{Astrophys. J.} \textbf{399}, 462

\bibitem[Bender et al.(1993)]{bend93} Bender, R., Burstein, D., Faber, S. M., Dynamically hot galaxies. I. Global stellar populations. 1993, \textit{Astrophys. J.} \textbf{411}, 153

\bibitem[Binney \& Tremaine(2007)]{binn07} Binney, J., Tremaine, S., ''Galactic Dynamics'', second eddition, Princeton University Press, 2007, US.

\bibitem[Borka et al.(2012)]{bork12} Borka, D., Jovanovi\'{c}, P., Borka Jovanovi\'{c}, V., Zakharov, A. F., Constraints on $R^n$ gravity from precession of orbits of S2-like stars. 2012, \textit{Phys. Rev. D} \textbf{85}, 124004

\bibitem[Borka et al.(2021)]{bork21} Borka, D., Borka Jovanovi\'{c}, V., Capozziello, S., Zakharov, A. F., Jovanovi\'{c}, P., Estimating the Parameters of Extended Gravity Theories with the Schwarzschild Precession of S2 Star. 2021, \textit{Universe} \textbf{7}, 407.

\bibitem[Borka Jovanovi\'{c} et al.(2016)]{bork16} Borka Jovanovi\'{c}, V., Capozziello, S., Jovanovi\'{c}, P., Borka, D., Recovering the fundamental plane of galaxies by $f(R)$ gravity. 2016, \textit{Phys. Dark Universe} \textbf{14}, 73

\bibitem[Borka Jovanovi\'{c} et al.(2019)]{bork19} Borka Jovanovi\'{c}, V., Jovanovi\'{c}, P., Borka, D., Capozziello, S., Fundamental plane of elliptical galaxies in $f(R)$ gravity: the role of luminosity. 2019, \textit{Atoms} \textbf{7}, 4

\bibitem[Burstein et al.(1997)]{burs97} Burstein, B., Bender, R., Faber, S. M., Nolthenius, R., Global relationships among the physical properties of stellar systems. 1997, \textit{Astron. J.} \textbf{114}, 1365

\bibitem[Busarello et al.(1997)]{busa97} Busarello, G., Capaccioli, M., Capozziello, S., Longo, G., Puddu, E., The relation between the virial theorem and the fundamental plane of elliptical galaxies. 1997, \textit{Astron. Astrophys.} \textbf{320}, 415

\bibitem[Capozziello et al.(2007)]{capo07} Capozziello, S., Cardone, V. F., Troisi, A., Low surface brightness galaxy rotation curves in the low energy limit of $R^n$ gravity: no need for dark matter? 2007, \textit{Mon. Not. R. Astron. Soc.} \textbf{375}, 1423

\bibitem[Capozziello \& De Laurentis(2011)]{capo11} Capozziello, S., De Laurentis, M., Extended theories of gravity. 2011, \textit{Phys. Rep.} \textbf{509}, 167

\bibitem[Capozziello et al.(2017)]{capo17} Capozziello, S., Jovanovi\'{c}, P., Borka Jovanovi\'{c}, V., Borka, D., Addressing the missing matter problem in galaxies through a new fundamental gravitational radius. 2017, \textit{J. Cosmol. Astropart. P.} \textbf{2017}, No. 06, 044

\bibitem[Capozziello et al.(2018)]{capo18} Capozziello, S., Borka, D., Borka Jovanovi\'{c}, V., Jovanovi\'{c}, P., Galactic Structures from Gravitational Radii. 2018, \textit{Galaxies} \textbf{6}, 22

\bibitem[Capozziello et al.(2020)]{capo20} Capozziello, S., Borka Jovanovi\'{c}, V., Borka, D., Jovanovi\'{c}, P., Constraining theories of gravity by fundamental plane of elliptical galaxies. 2020, \textit{Phys. Dark Universe} \textbf{29}, 100573

\bibitem[Cardone(2004)]{card04} Cardone, V. F., The lensing properties of the Sersic model. 2004, \textit{Astron. Astrophys.} \textbf{415}, 839

\bibitem[Cattaneo et al.(2014)]{catt14} Cattaneo A., Salucci, P., Papastergis, E., Galaxy luminosity function and Tully-Fisher relation: reconciled through rotation-curve studies. 2014, \textit{Astrophys. J.} \textbf{783}, 66

\bibitem[Ciotti(1996)]{ciot96} Ciotti, L., Lanzoni, B., Renzini, A., The tilt of the fundamental plane of elliptical galaxies - I. Exploring dynamical and structural effects. 1996, \textit{Mon. No. R. Astron. Soc.} \textbf{282}, 1

\bibitem[Fischbach(2004)]{fisc99} Fischbach, E., Talmadge, C. L., The Search for Non-Newtonian Gravity, Springer: Heidelberg, Germany; New York, NY, USA, 1999; 305p.

\bibitem[Lelli et~al.(2016)]{lell16} Lelli, F. et~al., The Small Scatter of the Baryonic Tully-Fisher Relation. 2016, \textit{Astrophys. J. Lett.} \textbf{816(1)}, L14

\bibitem[McGaugh et~al.(2000)]{mcga00} McGaugh, S. S. et~al., The Baryonic Tully-Fisher Relation. 2000, \textit{Astrophys. J. Lett.} \textbf{533}, L99

\bibitem[McGaugh(2011)]{mcga11} McGaugh, S. S., Novel Test of Modified Newtonian Dynamics with Gas Rich Galaxies. 2011, \textit{Phys. Rev. Lett.} \textbf{106}, 121303

\bibitem[McGaugh(2012)]{mcga12} McGaugh, S. S., The Baryonic Tully-Fisher Relation of Gas-rich Galaxies as a Test of $\Lambda$CDM and MOND. 2012, \textit{Astron. J.} \textbf{143(2)}, 40

\bibitem[Milgrom(1983)]{milg83} Milgrom, M., A modification of the Newtonian dynamics as a possible alternative to the hidden mass hypothesis. 1983, \textit{Astrophys. J.} \textbf{270}, 365
	
\bibitem[Nojiri \& Odintsov (2011)]{noji11} Nojiri, S., Odintsov, S. D., Unified cosmic history in modified gravity: from $F(R)$ theory to Lorentz non-invariant models. 2011, \textit{Phys. Rept.} \textbf{505}, 59

\bibitem[Nojiri et al.(2017)]{noji17} Nojiri, S., Odintsov, S. D., Oikonomou, V. K., Modified Gravity Theories on a Nutshell: Inflation, Bounce and Late-time Evolution. 2017, \textit{Phys. Rept.} \textbf{692}, 1

\bibitem[Said(2023)]{said23} Said, K., Tully-Fisher relation, 2023, arXiv:2310.16053

\bibitem[Sparke \& Gallagher(2007)]{spar07} Sparke, L. S., Gallagher, J. S., Galaxies in the Universe: An Introduction, Cambridge University Press, 2007, ISBN: 9780521671866

\bibitem[Tully \& Fisher(1977)]{tull77} Tully, R. B., Fisher, J. R., A new method of determining distances to galaxies. 1977, \textit{Astron. Astrophys.} \textbf{54}, 661

\bibitem[Yegorova et al.(2012)]{yego12} Yegorova, I. A., Babic, A., Salucci P., Spekkens, K., Pizzella, A., Rotation curves of luminous spiral galaxies. 2012, \textit{Astron. Astrophys. Trans.} \textbf{27}, 335

\bibitem[Zakharov et al.(2006)]{zakh06} Zakharov, A. F., Nucita, A. A., Paolis, F. D., Ingrosso, G., Solar system constraints on $R^n$ gravity. 2006, \textit{Phys. Rev. D} \textbf{74}, 107101

\bibitem[Zakharov et al.(2007)]{zakh07} Zakharov, A. F., Nucita, A. A., Paolis, F. D., Ingrosso, G., Apoastron shift constraints on dark matter distribution at the Galactic Center. 2007, \textit{Phys. Rev. D} \textbf{76}, 062001

\bibitem[Zakharov et al.(2014)]{zakh14} Zakharov, A. F., Borka, D., Borka Jovanovi\'{c}, V., Jovanovi\'{c}, P., Constraints on $R^n$ gravity from precession of orbits of S2-like stars: A case of a bulk distribution of mass. 2014, \textit{Adv. Space Res.} \textbf{54}, 1108

\end{thebibliography}

\end{document}